\documentclass[preprint2]{aastex}

\slugcomment{To appear in Astrophysical Journal Letters}

\shorttitle{Rotation Periods of Young Brown Dwarfs}
\shortauthors{Scholz et al.}

\begin{document}
\bibliographystyle{apj}


\title{Rotation Periods of Young Brown Dwarfs: \\K2 Survey in Upper Scorpius}

\author{Alexander Scholz\altaffilmark{1}, Veselin Kostov\altaffilmark{2}, Ray Jayawardhana\altaffilmark{3}, Koraljka Mu\v{z}i\'c\altaffilmark{4,5}}

\email{as110@st-andrews.ac.uk}

\altaffiltext{1}{School of Physics and Astronomy, University of St Andrews, North Haugh, St Andrews, Fife KY16 9SS, United Kingdom}
\altaffiltext{2}{Department of Astronomy \& Astrophysics, University of Toronto, 50 St. George Street, Toronto, ON M5S 3H4, Canada}
\altaffiltext{3}{Faculty of Science, York University, 355 Lumbers Building, 4700 Keele Street, Toronto, ON M3J 1P2, Canada}
\altaffiltext{4}{Nucleo de Astronom\'ia, Facultad de Ingenier\'ia, Universidad Diego Portales, Av. Ejercito 441, Santiago, Chile}
\altaffiltext{5}{European Southern Observatory, Alonso de C\'ordova 3107, Casilla 19, Santiago 19001, Chile}

\begin{abstract}
We report rotational periods for 16 young brown dwarfs in the nearby Upper Scorpius association, based on 72 days of high-cadence, 
high-precision photometry from the {\it Kepler} space telescope's K2 mission. The periods range from a few hours to two days (plus one
outlier at 5\,days), with a median just above one day, confirming that brown dwarfs, except at the very youngest ages, are fast rotators. 
Interestingly, four of the slowest rotators in our sample exhibit mid-infrared excess emission from disks; at least two also show signs 
of disk eclipses and accretion in the lightcurves. Comparing these new periods with those for two other young clusters and simple 
angular momentum evolution tracks, we find little or no rotational braking in brown dwarfs between 1-10 Myr, in contrast to low-mass 
stars. Our findings show that disk braking, while still at work, is inefficient in the substellar regime, thus provide an important 
constraint on the mass dependence of the braking mechanism.
\end{abstract}

\keywords{}

\section{Introduction}
\label{s1}

Rotation is a key parameter in stellar evolution, and is directly linked to many fundamental astrophysical processes, 
including magnetic field generation, stellar winds, star-disk interaction, and binary formation \citep{2007prpl.conf..297H,2014prpl.conf..433B}. 
Because rotation drives magnetic activity, it is also relevant for 'space weather' in the stellar 
surroundings and thus for planetary habitability \citep{2007LRSP....4....3G,2014MNRAS.438.1162V}. The rotation period, 
derived from the periodic photometric modulation induced by surface spots, is one of the rare stellar parameters 
that can be measured with high precision ($\sim 1$\%) for large samples.

Large numbers of periods have been measured for low-mass stars with spectral types F to M of all ages from $\sim 1$\,Myr 
to several Gyrs \citep[e.g.][]{2002A&A...396..513H,2005A&A...430.1005L,2008MNRAS.383.1588I,2014ApJS..211...24M}, and they provide
very useful constraints for angular momentum evolution models \citep[e.g.][]{2015A&A...577A..98G}. Other portions 
of the mass-age parameter space are still poorly explored. One example is the regime of BDs, substellar objects 
with masses less than $\sim 0.08\,M_{\odot}$. The only regions with significant samples of 
measured BD periods are clusters in Orion, the ONC \citep{2009A&A...502..883R} $\sigma$\,Ori 
\citep{2004A&A...419..249S,2010ApJS..191..389C}, 
and $\epsilon$\,Ori \citep{2005A&A...429.1007S}, at ages of 1-5\,Myr. Like stars, BDs start with a range of periods 
from hours up to 5-10 days, and spin up due to pre-main sequence contraction, but in contrast to stars they maintain fast 
rotation with negligible spindown for several Gyrs. 

Due to their intrinsic faintness, the small photometric amplitudes, and the wide range of periods at young ages, it is 
challenging to obtain a reliable picture of the rotational evolution of BDs with ground-based observations alone, which feature 
typical daytime gaps and are therefore biased in the period sensitivity. There are good indications that disk braking 
is less efficient in the very low mass regime \citep[e.g.][]{{2005A&A...430.1005L}}. But the lack of substantial and unbiased 
period samples for a range of ages prohibits us from putting firm constraints on models.

Here we present first period measurements for BDs obtained with Kepler's K2 mission \citep{2014PASP..126..398H}. In campaign 2, 
the K2 field included large parts of the Upper Scorpius (UpSco) star forming region (age 5-10 Myr), which harbours a huge 
population of BDs \citep[e.g.][]{2008MNRAS.383.1385L,2008ApJ...688..377S,2011A&A...527A..24L,2014MNRAS.442.1586D}. 
With a cadence of 30\,min and a time baseline of 72\,d, the K2 lightcurves cover the entire period range for young BDs. 
During the course of its mission, K2 is expected to fill the period-age diagram for BDs with dozens of new 
datapoints; this study is the first step towards a new census of BD rotation.

\section{Data processing}
\label{s2}

We observed brown dwarfs\footnote{Some of our objects may be very low mass stars with masses slightly above the substellar
limit; for convenience we will call all targets brown dwarfs.} in UpSco as part of campaign 2 of the K2 mission. The
targets were selected from the census by \citet{2013MNRAS.429..903D} based on the UKIDSS survey. Analysis of mid-infrared data
shows that $\sim 20$\% still harbour circum-sub-stellar disks. 51 of the full sample of 116 have been covered by K2.

We downloaded the data, in the form of long-cadence target pixel files, from the Mikulsi Archive for Space Telescopes (MAST). 
To extract aperture photometry, detrend and correct for systematic effects we used the latest version of PyKE \citep{2012ascl.soft08004S}. 
This is an open-source PyRAF package originally designed for custom manipulation of the {\it Kepler} data and now also 
including the self-flat-fielding procedure of \citet{2014PASP..126..948V} optimized for processing data from the K2 mission. 
The procedure utilizes the relation between the position of the target on the detector and the measured flux to correct for 
artifacts induced by the pointing drift of the spacecraft. 

\begin{figure*}
\center
\includegraphics[width=16cm]{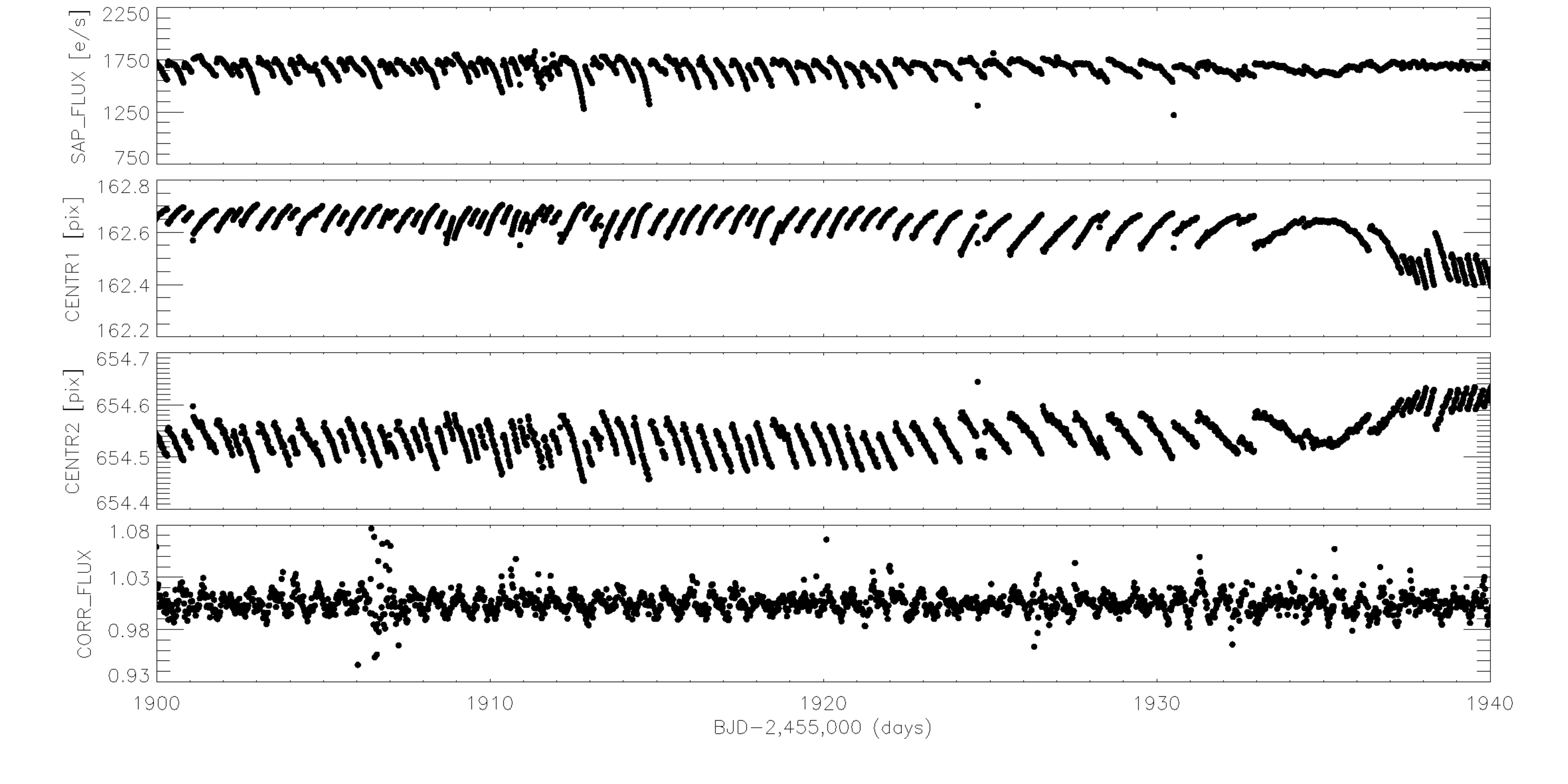} 
\caption{Processing of the SAP lightcurve for EPIC204614722, see text for details.
\label{proc}}
\end{figure*}

Our data reduction proceeds as follows. First we used PyKE's {\it kepmask} to define sub-image apertures ($\sim 2-3 \times 2-3$ pixels) 
from the respective target's postage stamp image (typically $10-12 \times 10-12$ pixels), then extract simple aperture photometry 
(SAP) with {\it kepextract}. Next, we detrend and normalize the raw lightcurves with {\it kepflatten} using a low-order polynomial 
(n~$\sim3-5$) and, finally, to correct for the motion-induced systematic effects we use {\it kepsff}. As an illustrative example, 
on Fig. \ref{proc} we show a section of the SAP lightcurve (background-corrected) of EPIC204614722 (top panel), the 
two moment centroids (middle panels) and the corresponding {\it kepsff}-corrected lightcurve (lower panel). Throughout the 
data-reduction process we appropriately adjust the relevant input parameters on a target-by-target basis.  We note that we used 
the entire campaign 2 dataset, including the majority of cadences from the second half of the campaign where most of the data 
flags are non-zero due to the photometer's local detector electronics parity errors, as these do not affect the quality of the 
data for the purposes of detecting rotational variability. 

\section{Lightcurve analysis}

All 51 lightcurves were visually inspected. 16 of them show clear signs of periodic variability. In most cases the periodicity
is obvious and periods ranging from a few hours up to 5\,d are estimated by eye. To measure the periods, we used the autocorrelation 
function, which has been established as a useful tool for period analysis in Kepler lightcurves by \citet{2014ApJS..211...24M}. 
Compared to Fourier (Lomb-Scargle) or phase-based techniques (e.g., phase dispersion minimisation), it has the advantage that it is 
robust against phase/amplitude changes as well as systematics. 

\begin{figure*}
\center
\includegraphics[width=4.0cm]{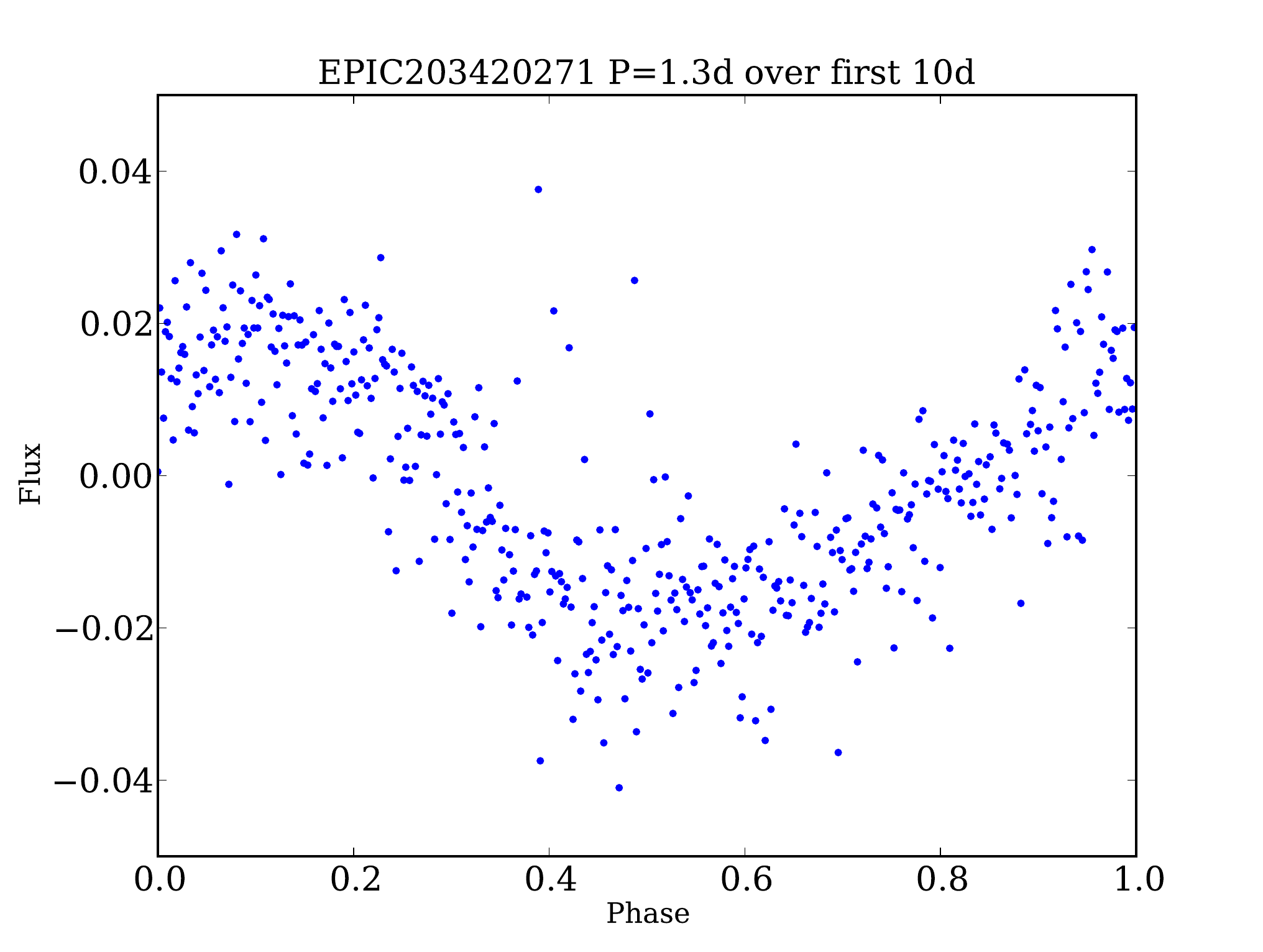} 
\includegraphics[width=4.0cm]{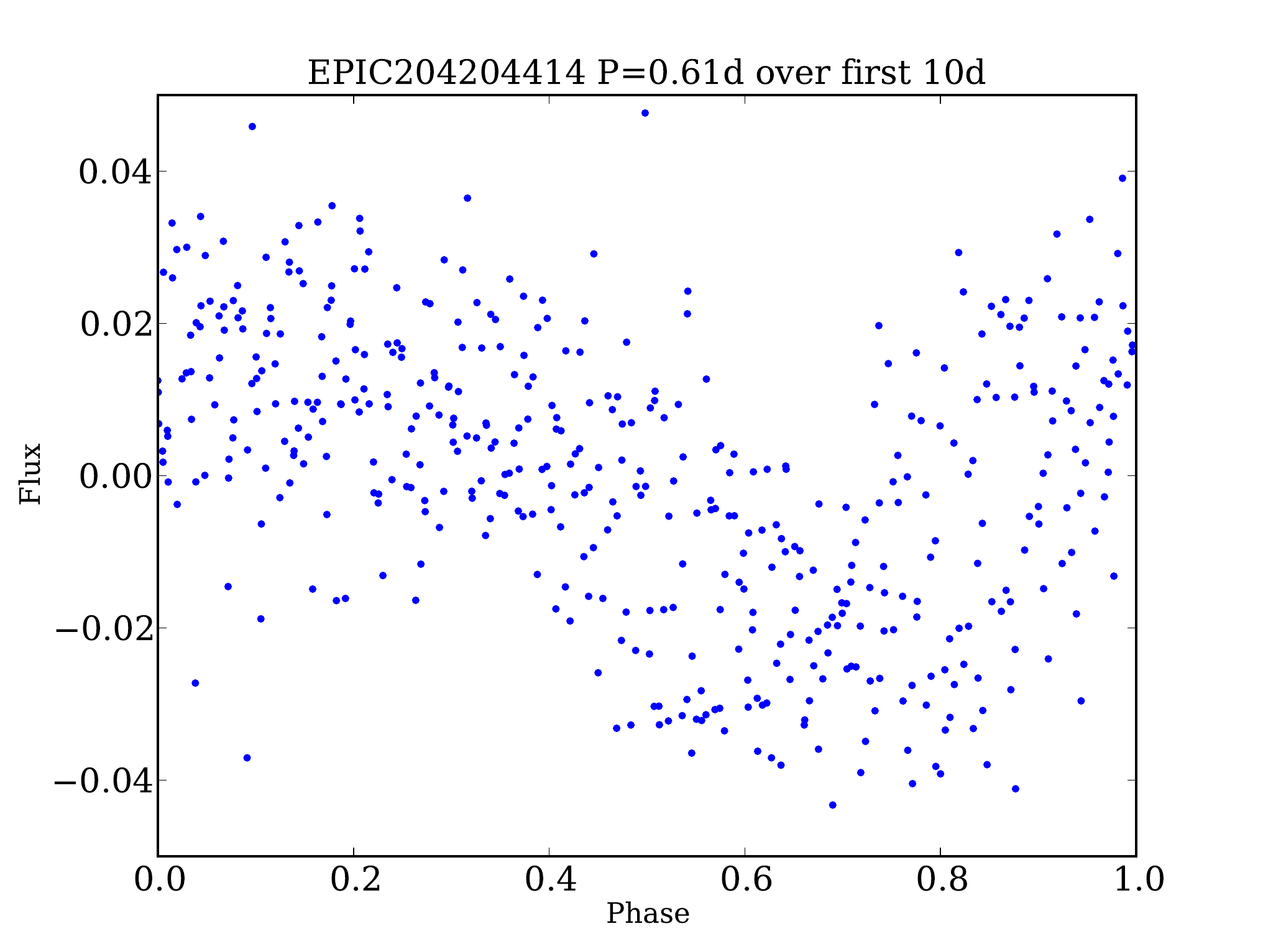}
\includegraphics[width=4.0cm]{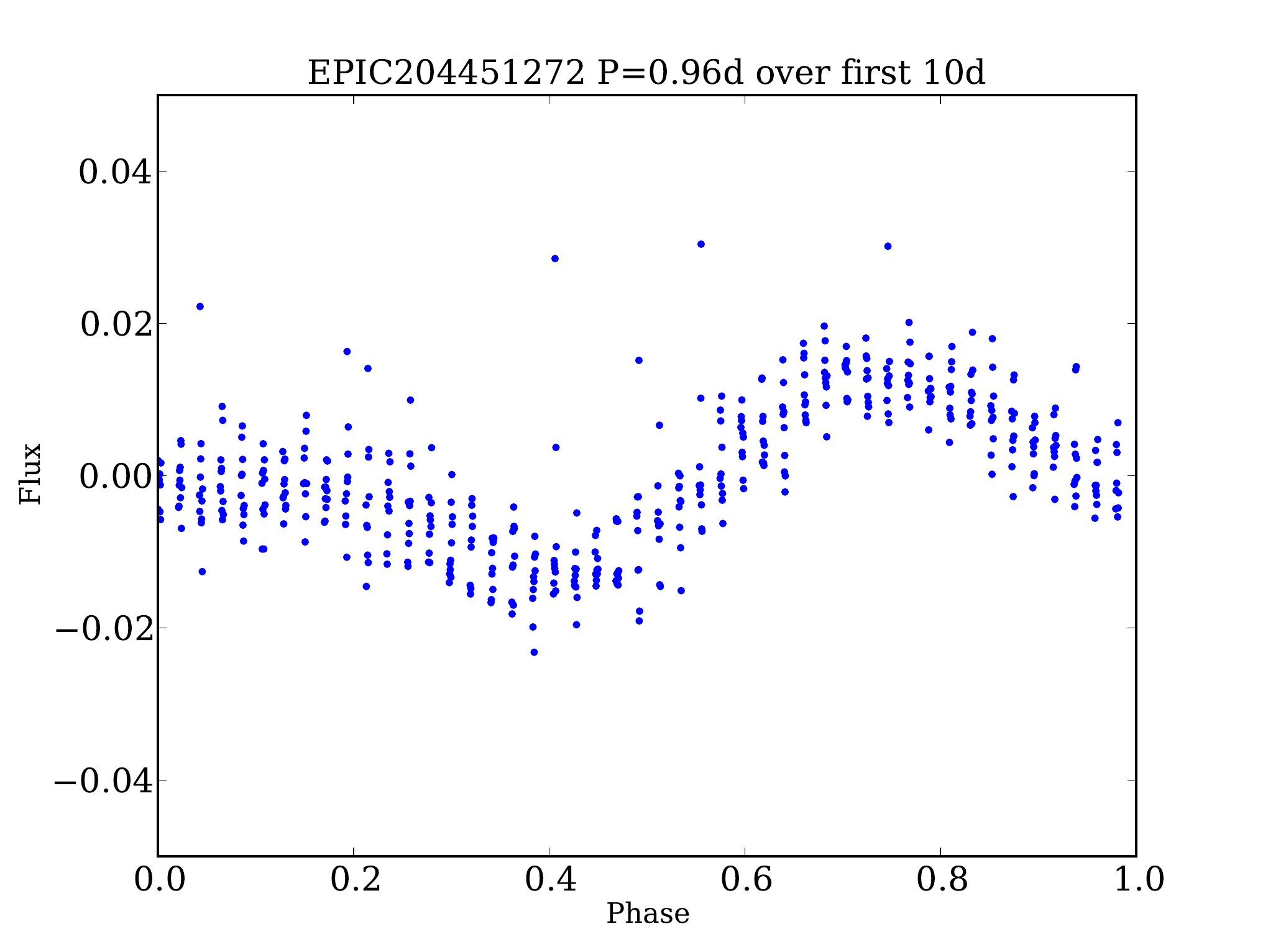} 
\includegraphics[width=4.0cm]{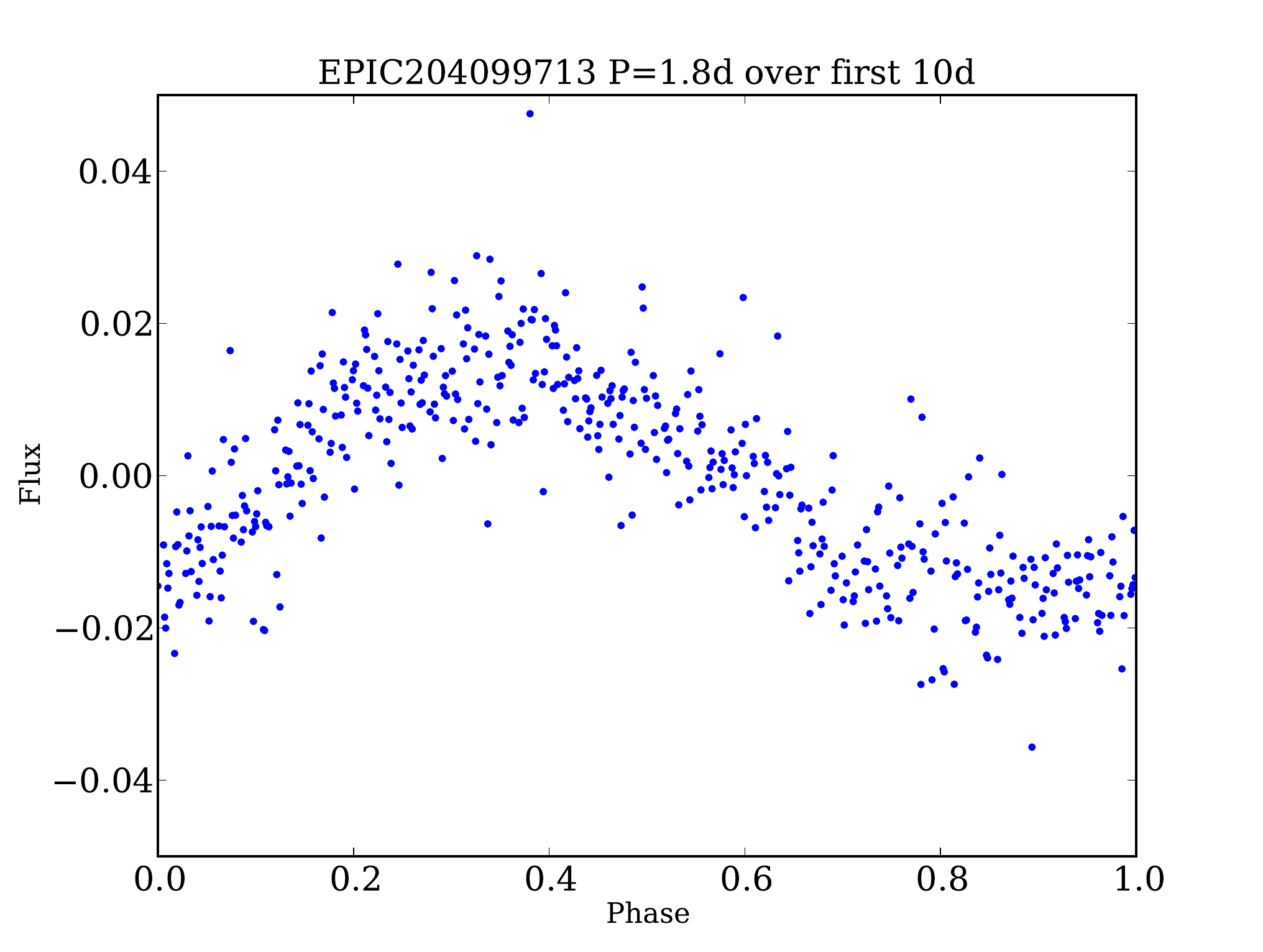} \\
\includegraphics[width=4.0cm]{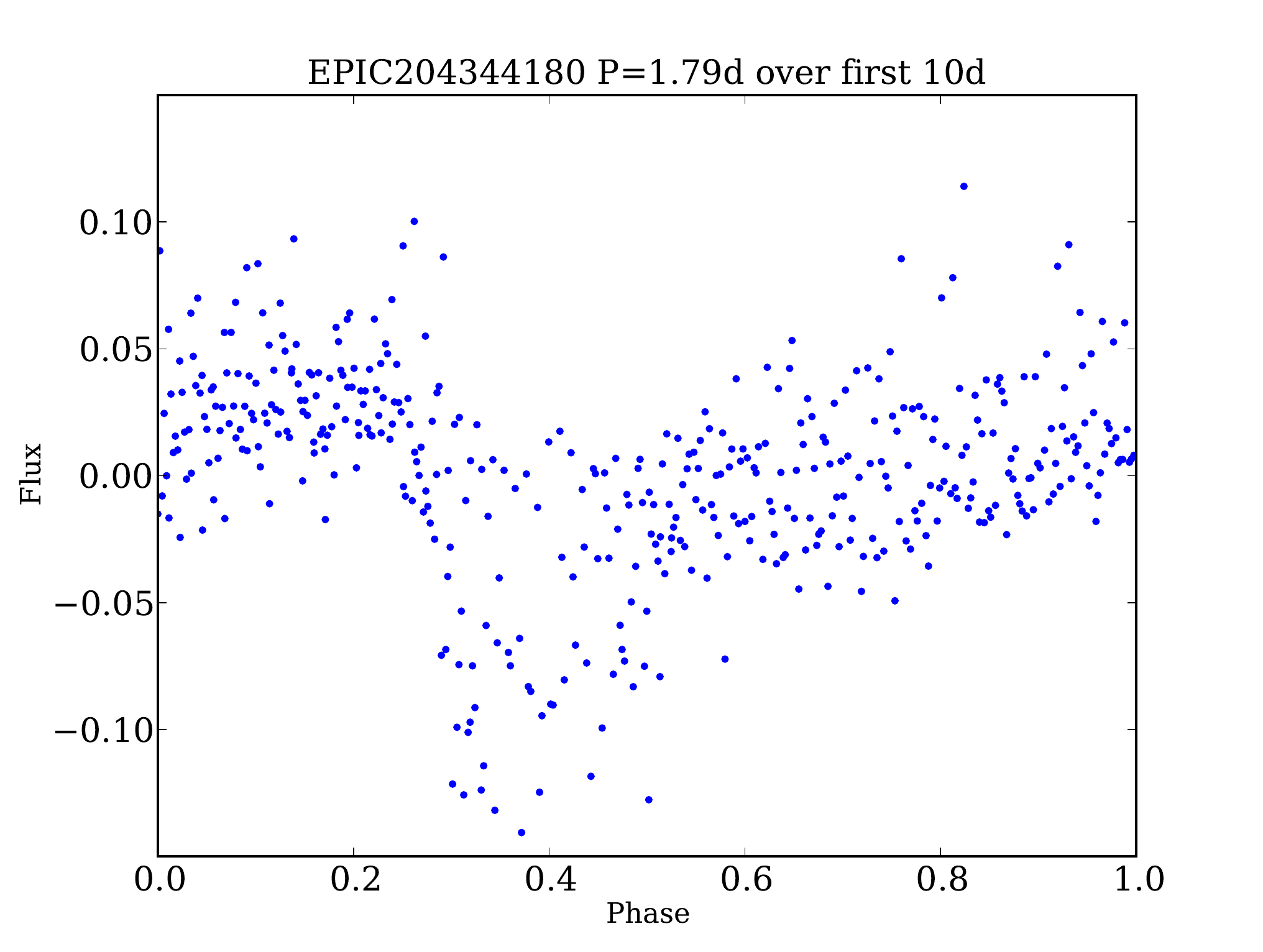} 
\includegraphics[width=4.0cm]{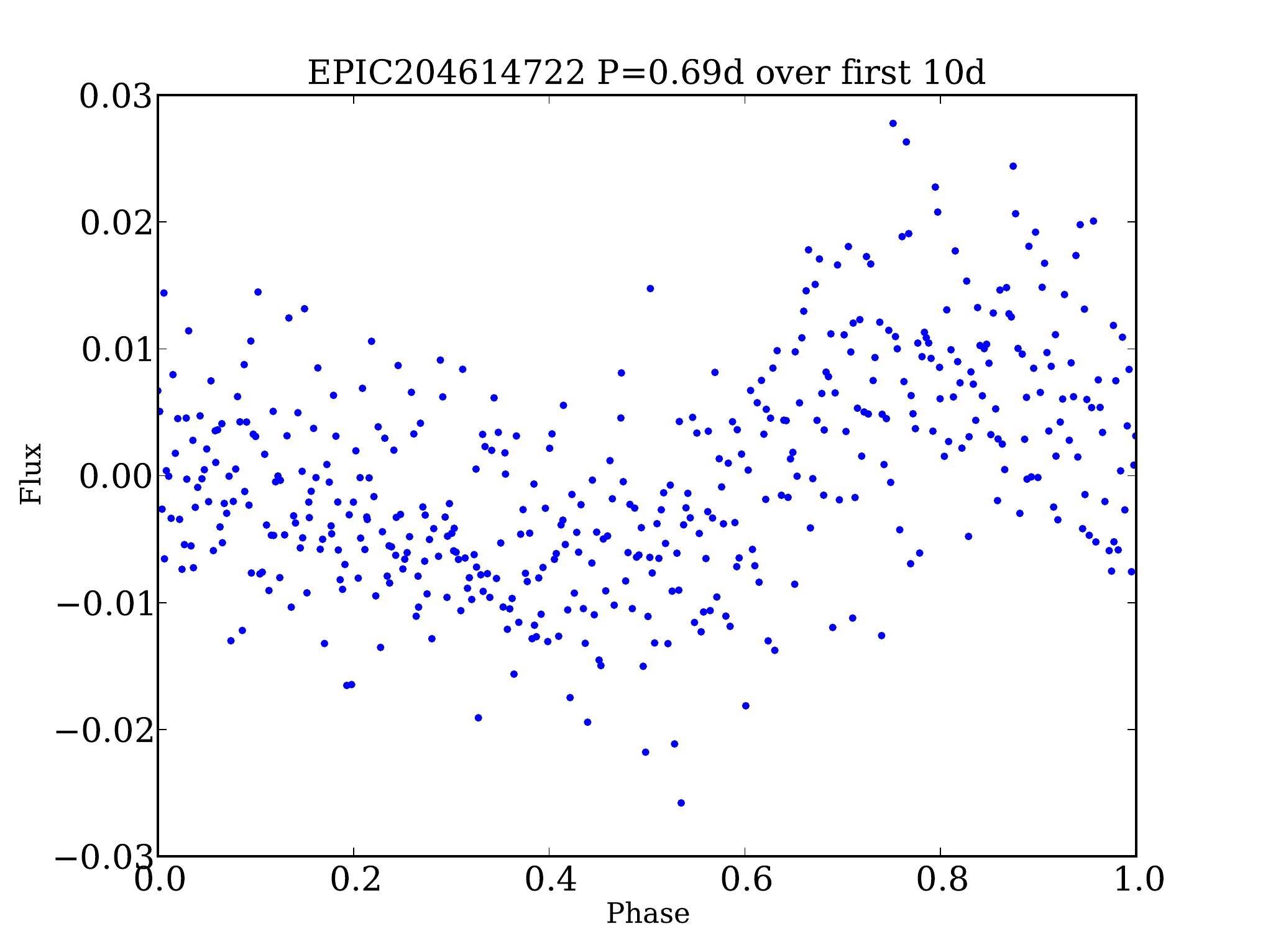} 
\includegraphics[width=4.0cm]{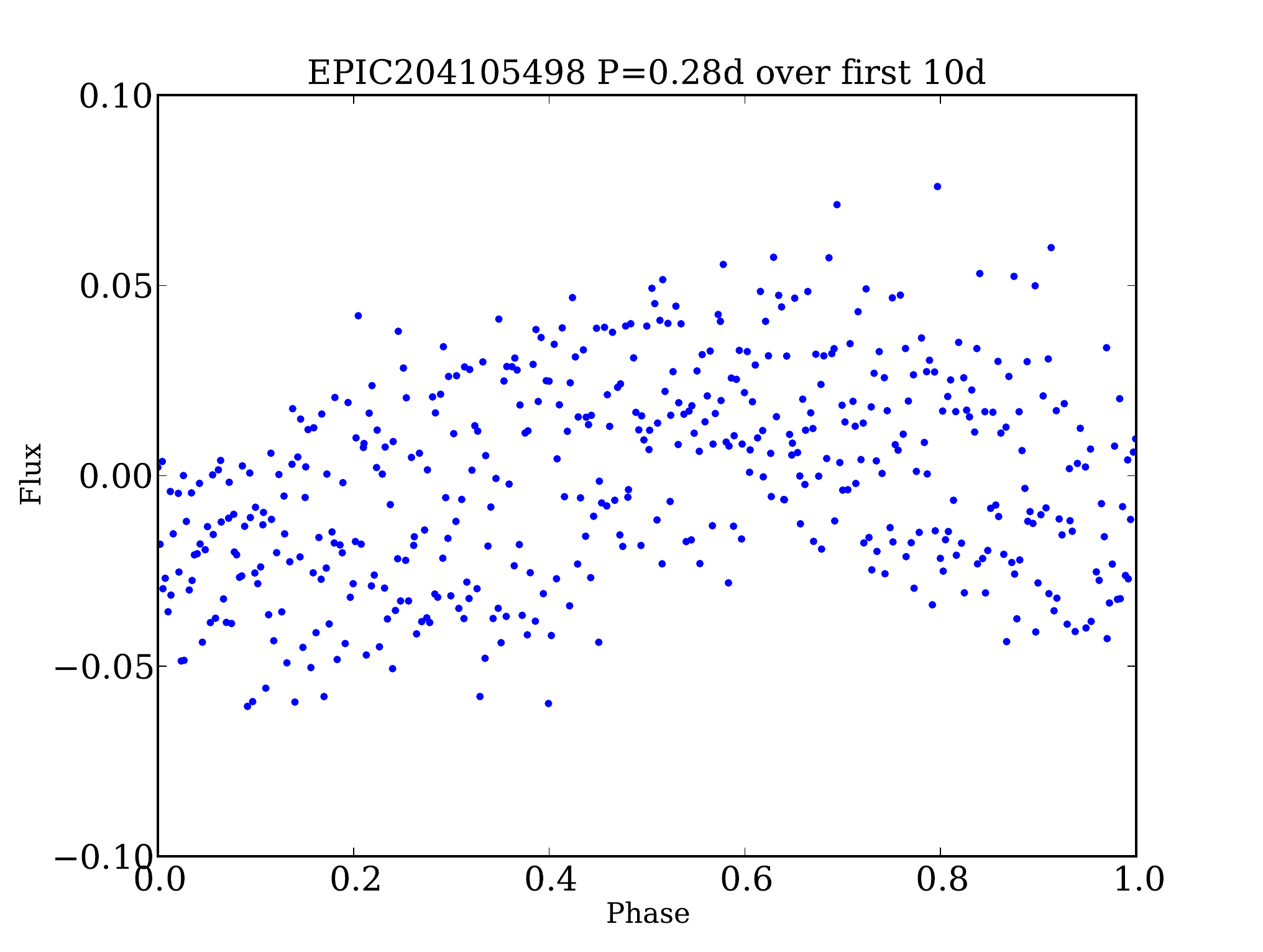} 
\includegraphics[width=4.0cm]{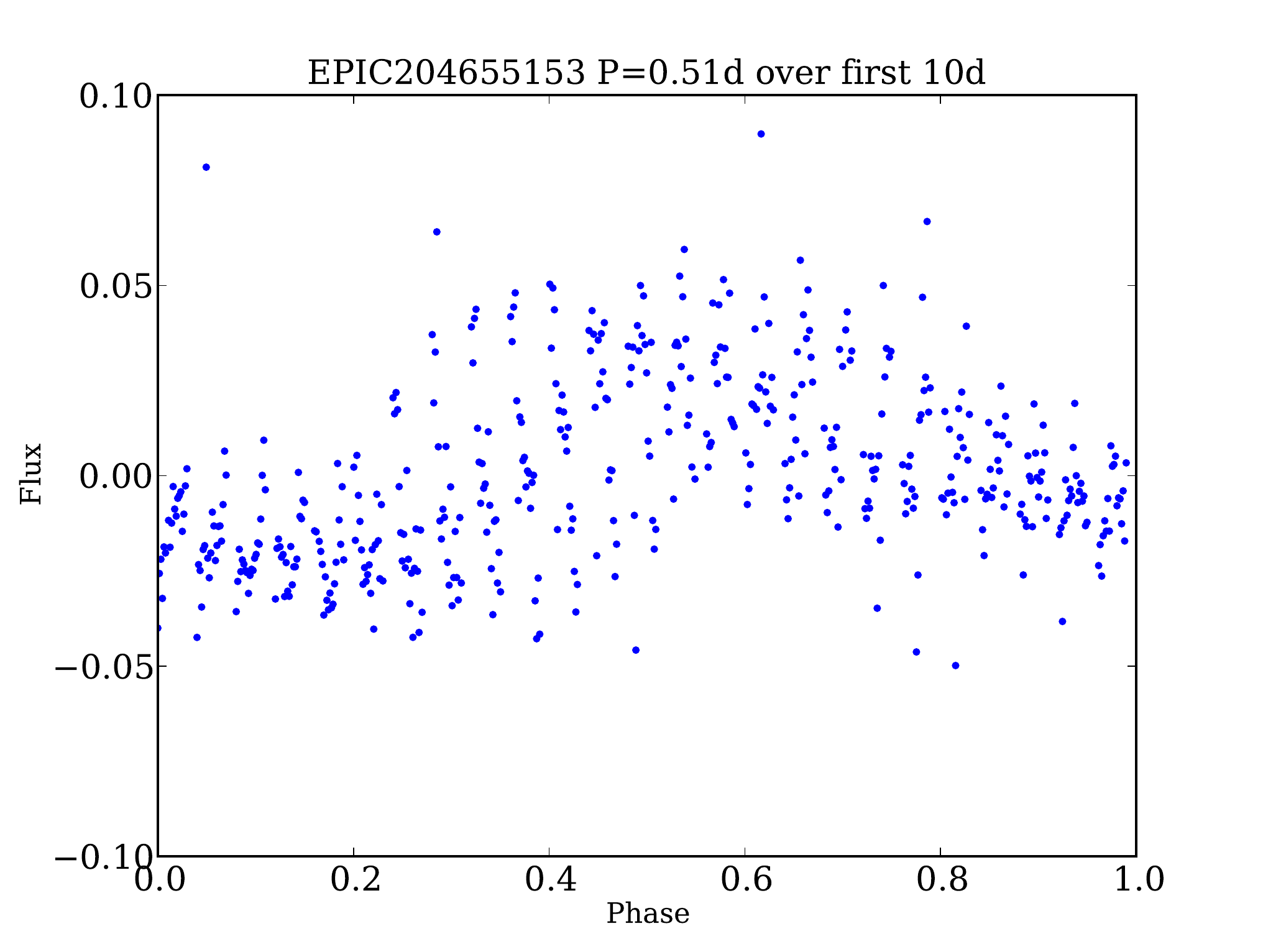} \\
\includegraphics[width=4.0cm]{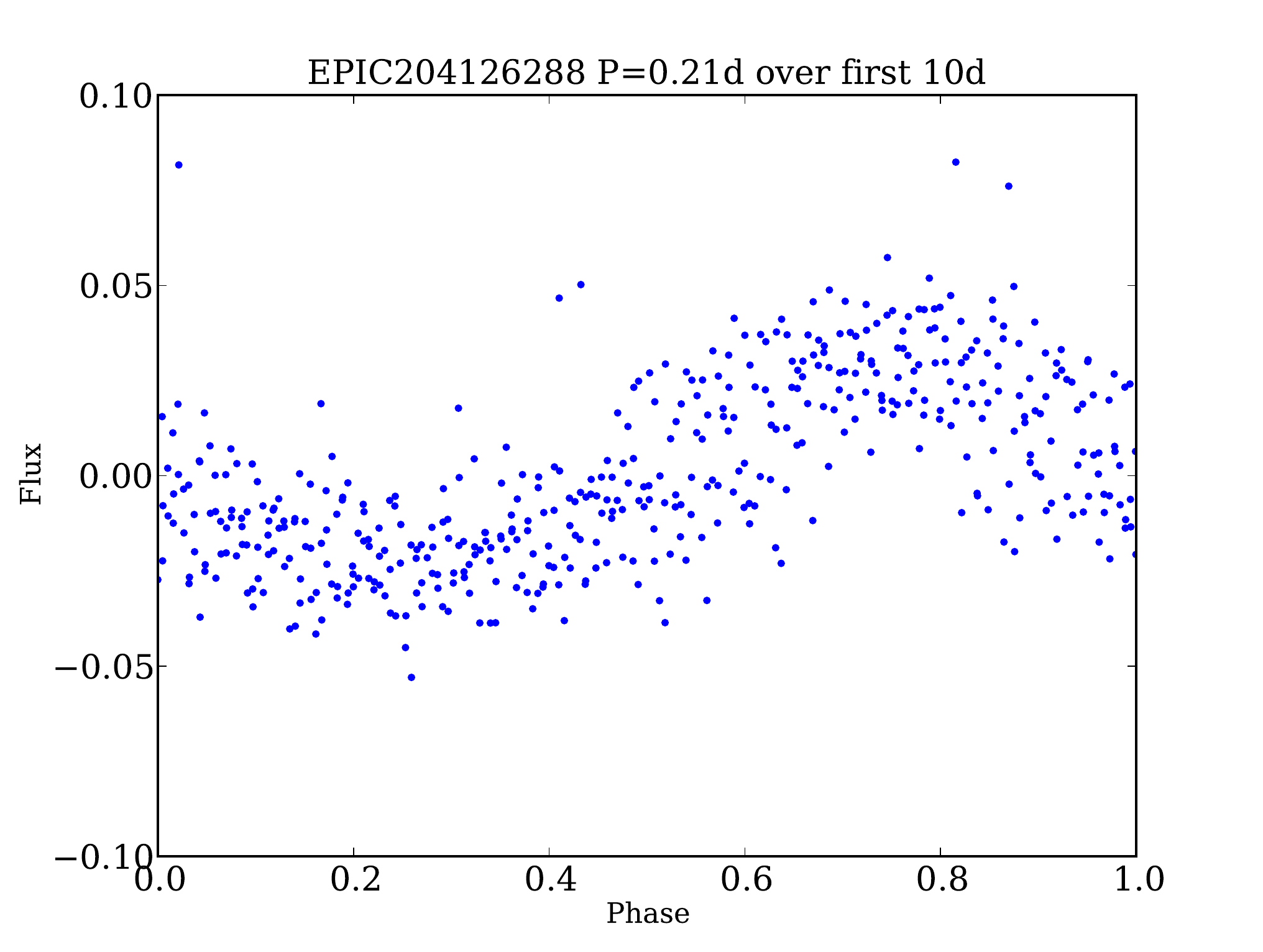} 
\includegraphics[width=4.0cm]{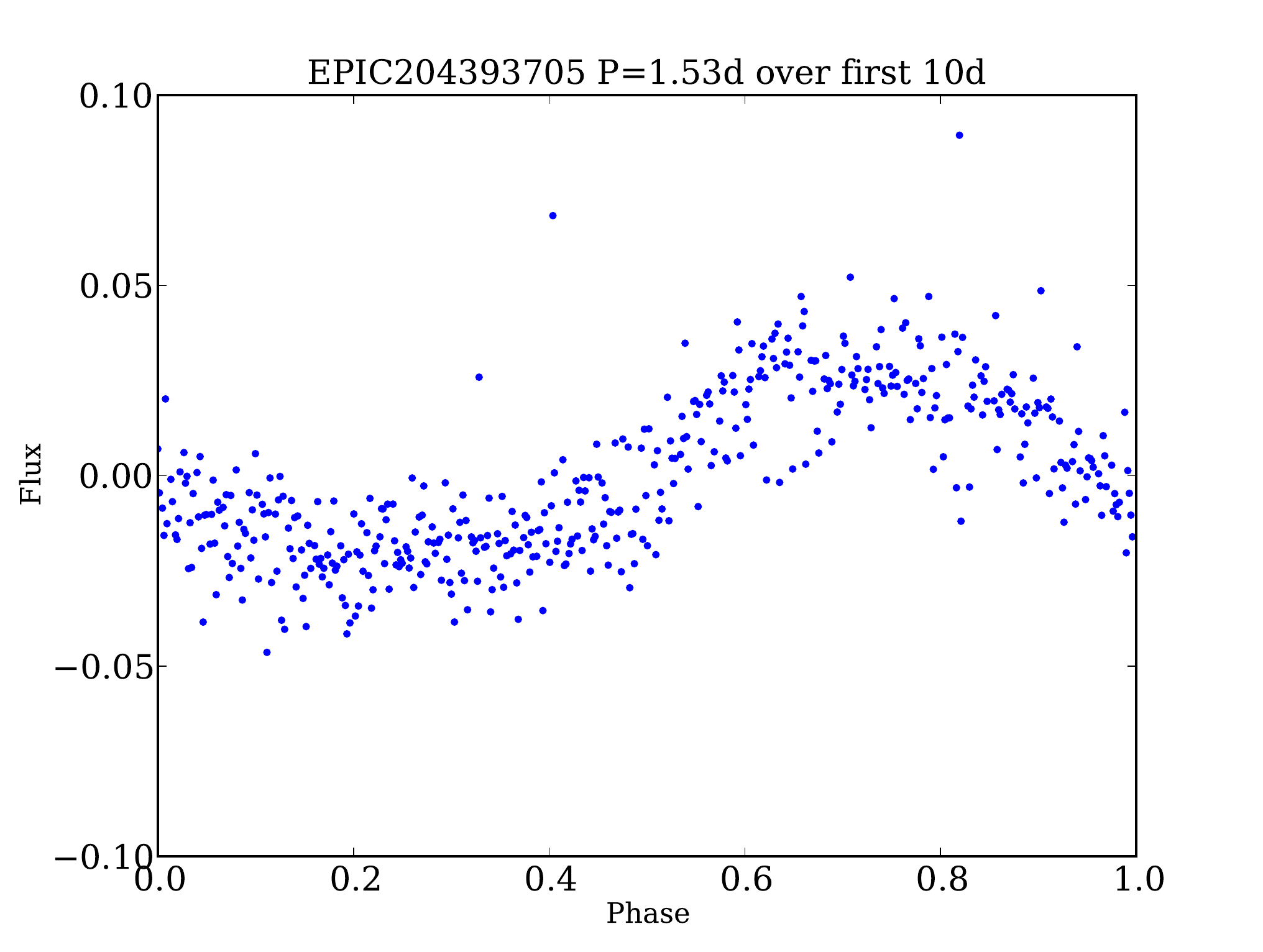} 
\includegraphics[width=4.0cm]{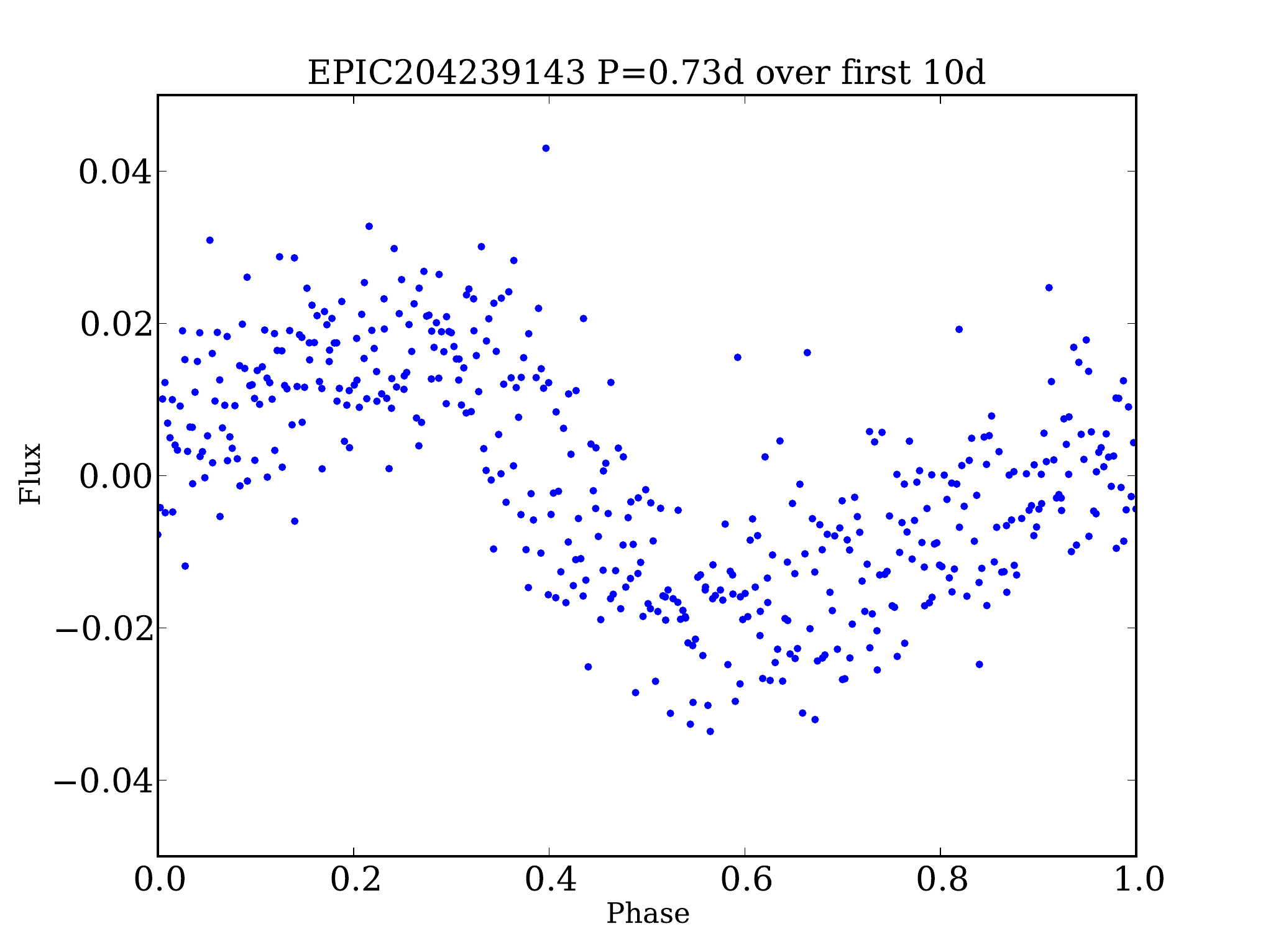} 
\includegraphics[width=4.0cm]{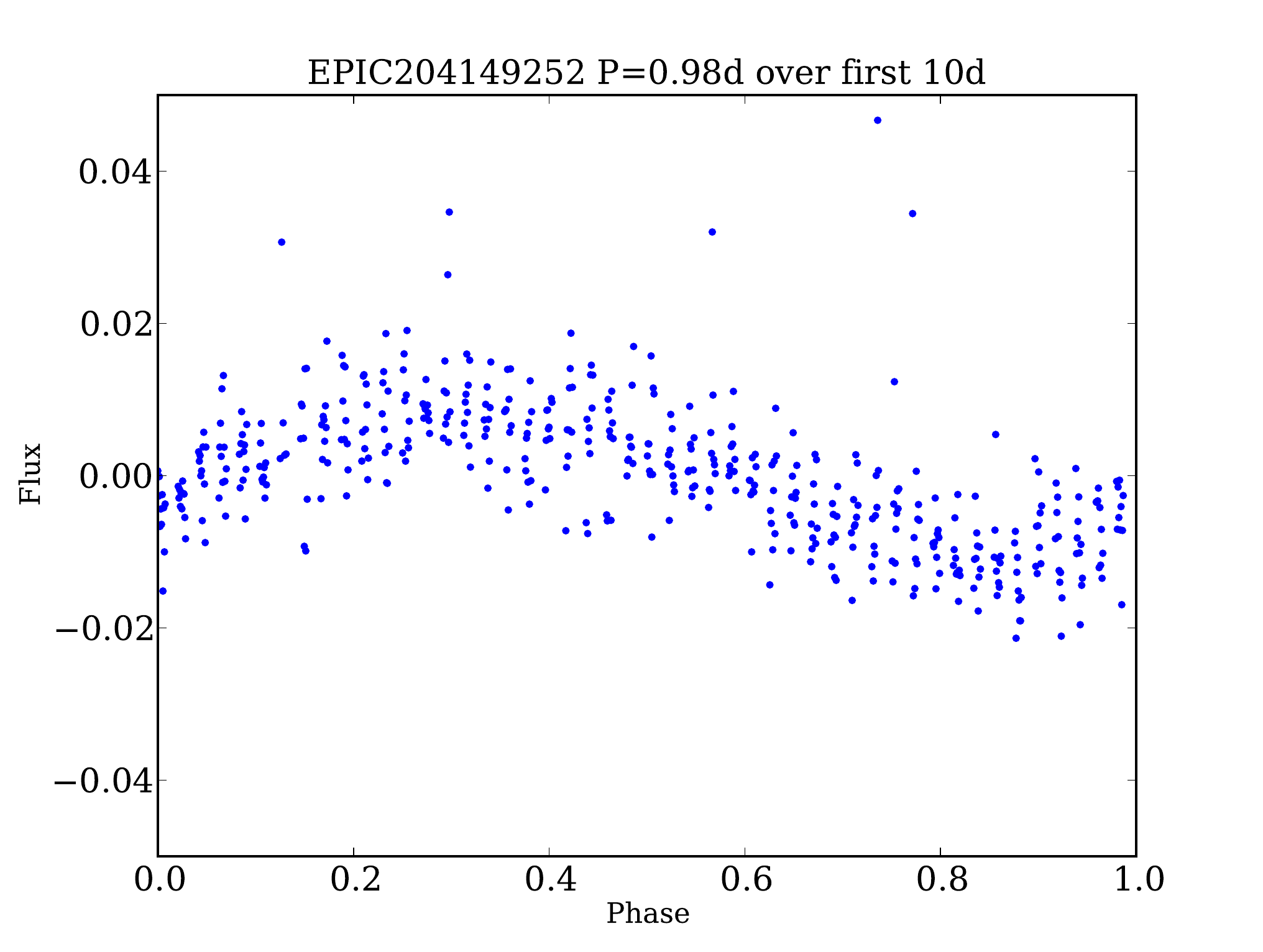} \\
\includegraphics[width=4.0cm]{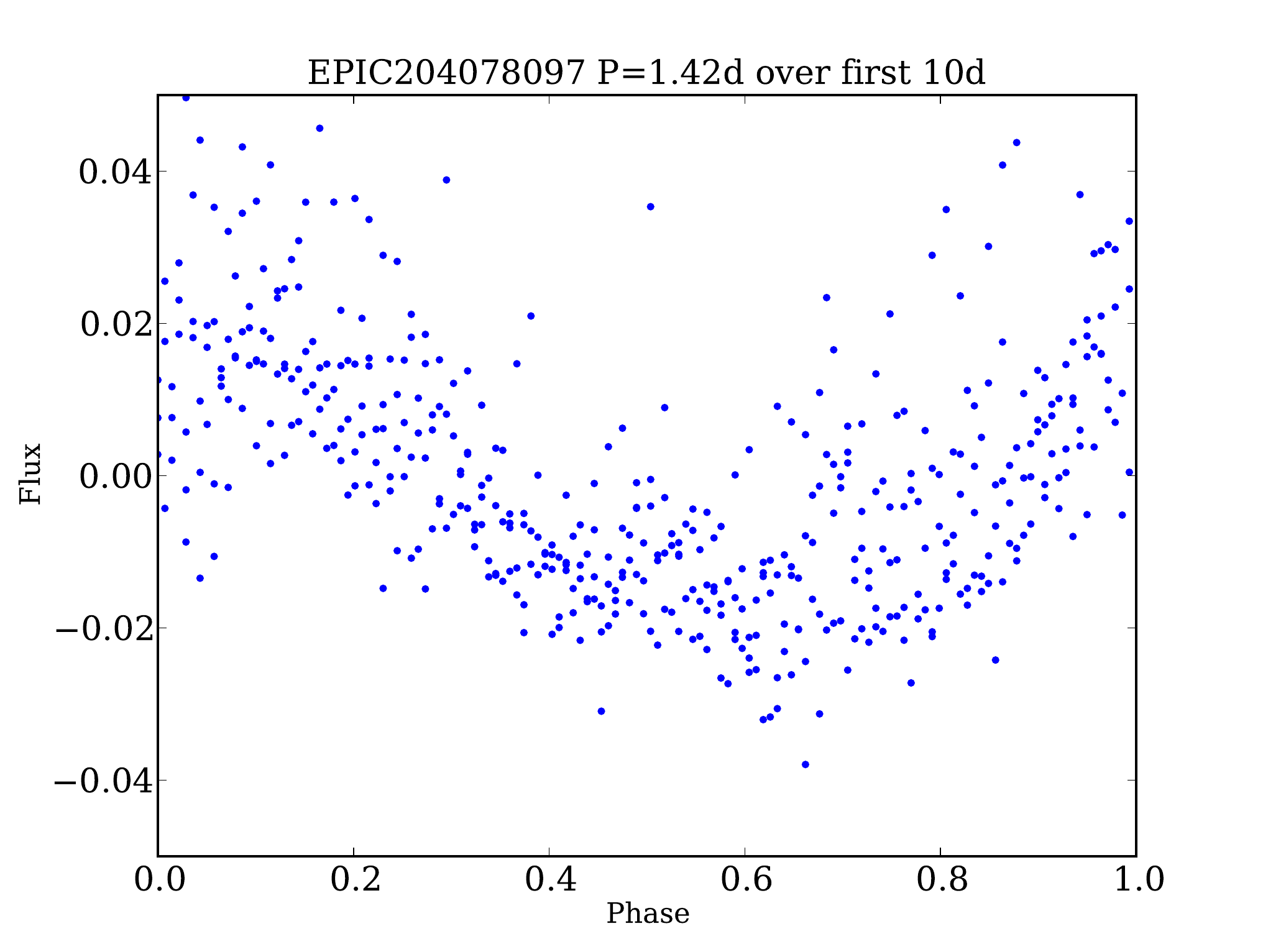} 
\includegraphics[width=4.0cm]{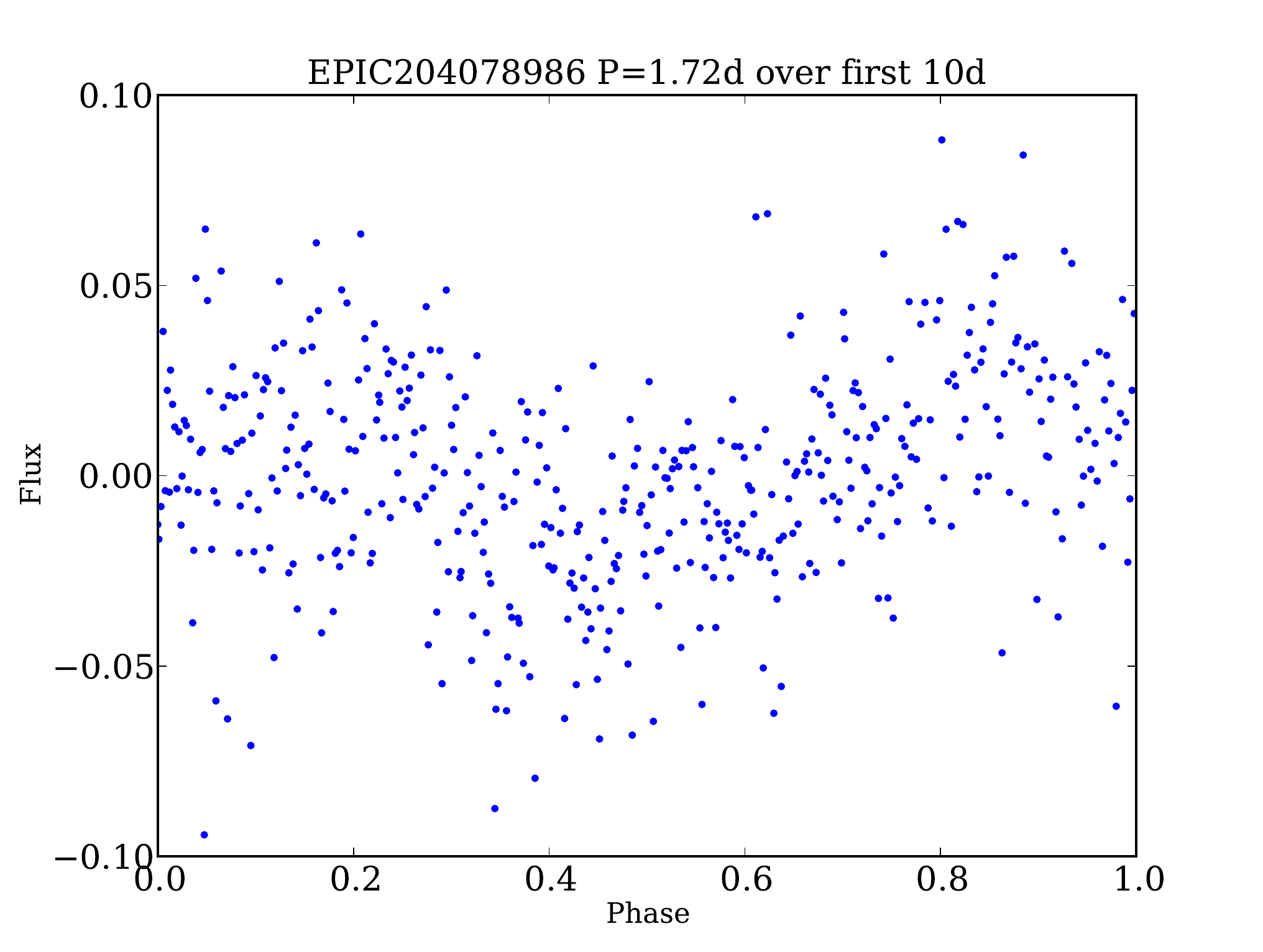} 
\includegraphics[width=4.0cm]{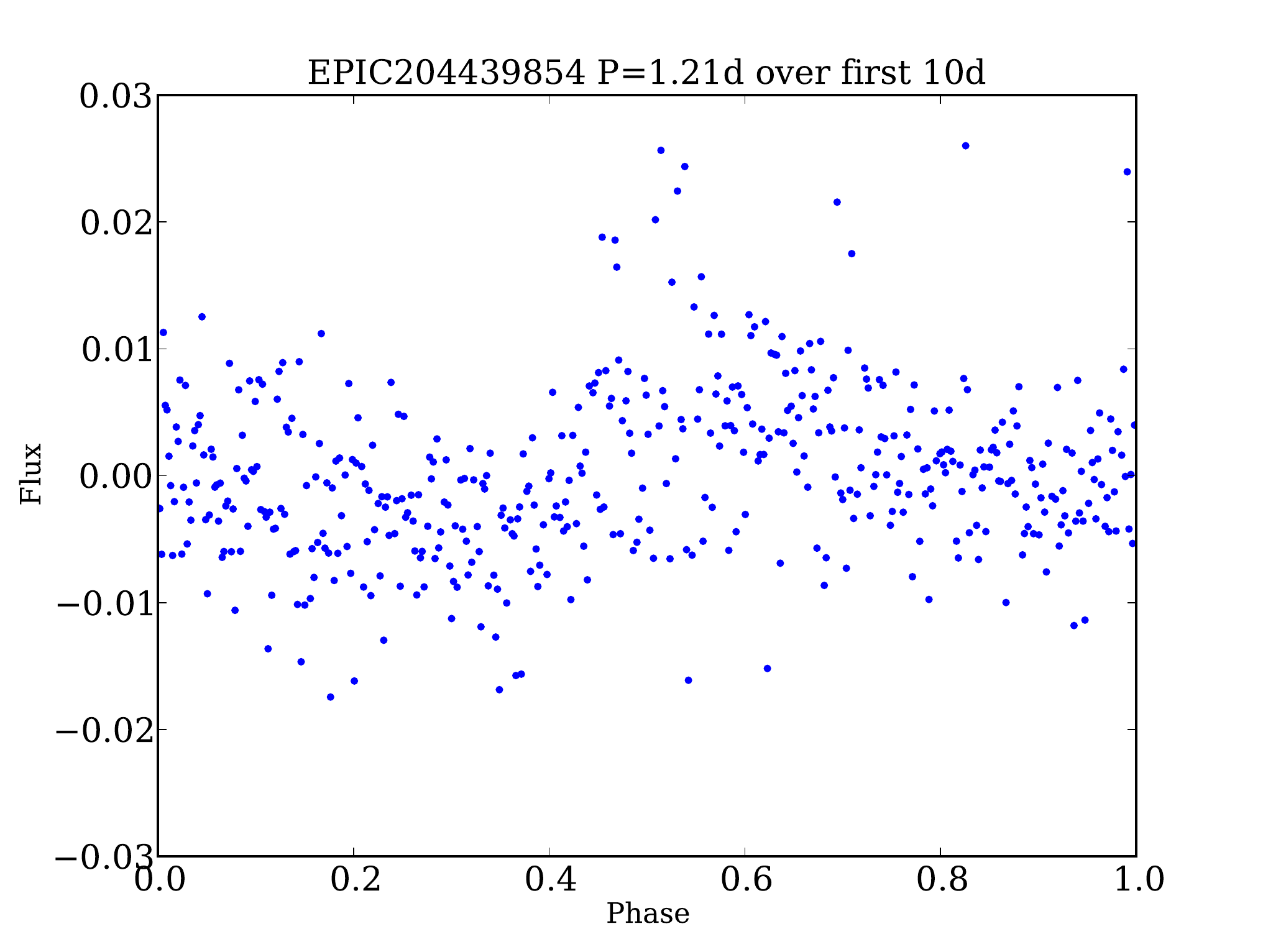} 
\includegraphics[width=4.0cm]{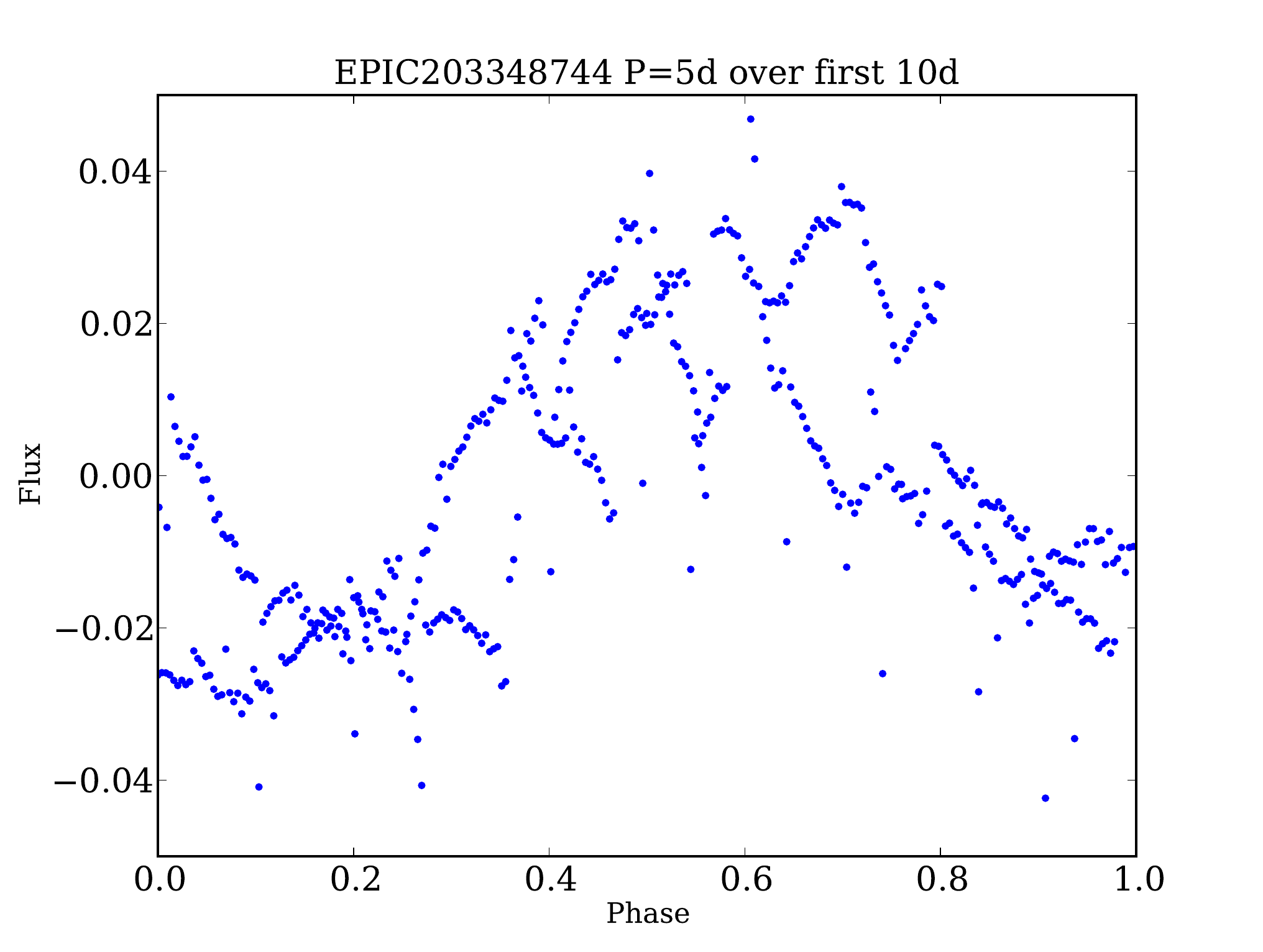} 
\caption{Phased lightcurves for the first ten days of the K2 campaign 2 for the 16
objects with detected periods, in the same order as in Table \ref{t1}.
\label{phase1}}
\end{figure*}

After dividing the lightcurves in seven segments of 500 datapoints each, each covering about 10\,d with a uniform cadence of 30\,min, 
we calculate the autocorrelation function and determine the position of the first non-zero peak, which is a measurement of the fundamental 
period in the dataset. For 10 objects, we find a consistent period in all 7 segments, for 5 more the same period is 
found for 5 or 6 segments. The scatter in the individual period measurements is 1-8\%, which we adopt as the uncertainty. One 
additional object shows an obvious $\sim 5$\,d period, but also strong additional variations, which could be intrinsic to the object,
but might also be partly instrumental. 
The best periods and the uncertainties are listed in Table \ref{t1}. The lightcurves are plotted in phase to the best period
(for the first 10 days of the campaign) in Fig. \ref{phase1}.\footnote{We note that we find a period of 0.48$\,d$ for 
three targets not in Table \ref{t1}, EPIC204367193, EPIC204418005, and EPIC204086791. We discard these because the lightcurves, phase plots 
and autocorrelation functions do not look convincing. These could be residuals of the systematics and have to be re-examined when
new software is available.}

To check for false positive signals due to contamination from other sources, we extracted the light curves of K2 stars near our targets as
listed in MAST (typically much brigher, and within $\sim2-3\arcsec$). We also tested various apertures on a target-by-target basis -- from the
central pixels, from the pixels surrounding them, and from any other conspicuously bright pixel in their respective postage stamp images -- and
visually examined the extracted light curves. Our analysis indicates that the flux modulations in our targets' light curves originate in their
respective central pixels only.  In addition, with the exception of Mars moving across the detector containing our targets for a few days in
October of 2014, there are no obvious false positive sources. We also confirmed that the periods estimated here are also visible in the
lightcurves made publicly available by \citet{2014PASP..126..948V}\footnote{{\tt https://www.cfa.harvard.edu/~avanderb/k2.html}}. Therefore we 
are convinced that the periods in Table \ref{t1} are of astrophysical origin.

\begin{table*}
\caption{UpSco rotation periods. \label{t1}}
\begin{tabular}{llllcc}
\tableline\tableline
EPIC & 2MASS & Kmag & $T_{\mathrm{eff}}$ (K)\tablenotemark{d} & P (d) & Perr (d) \\
\tableline
203420271 & 16033799-2611544 & 15.167 &  	  & 1.30 & 0.01 \\ 
204204414 & 16153648-2315175 & 16.417 & 3025	  & 0.61 & 0.02 \\ 
204451272 & 16082229-2217029 & 15.556 & 3025	  & 0.96 & 0.02 \\ 
204099713 & 16112630-2340059 & 15.971 & 2818, 3025 & 1.80 & 0.03 \\ 
204344180 & 16143287-2242133 & 16.775 & 2630, 2935 & 1.79 & 0.06\tablenotemark{a,b}\\ 
204614722 & 16095217-2136277 & 13.933 & 2570	  & 0.66 & 0.03 \\ 
204105498 & 16124692-2338408 & 14.964 & 2754, 2910 & 0.28 & 0.02 \\ 
204655153 & 16105499-2126139 & 15.316 & 2754	  & 0.51 & 0.01 \\ 
204126288 & 16164539-2333413 & 16.293 & 2884, 3025 & 0.21 & 0.01 \\ 
204393705 & 16132665-2230348 & 16.123 & 2910	  & 1.53 & 0.03 \\ 
204239143 & 16113837-2307072 & 15.143 & 2910	  & 0.73 & 0.02 \\ 
204149252 & 16133476-2328156 & 14.843 & 3025	  & 0.98 & 0.02 \\ 
204078097 & 16095852-2345186 & 13.992 & 2935	  & 1.42 & 0.04\tablenotemark{a,c} \\ 
204078986 & 16080745-2345055 & 15.786 & 2935	  & 1.72 & 0.08\tablenotemark{a} \\ 
204439854 & 16113470-2219442 & 14.662 & 2035	  & 1.21 & 0.05 \\ 
203348744 & 16030235-2626163 & 14.614 & & $\sim 5$ & 0.5\tablenotemark{a} \\ 
\tableline
\end{tabular}

\tablenotetext{a}{mid-infrared excess emission}
\tablenotetext{b}{irregular eclipses in the first part of the lightcurve}
\tablenotetext{c}{rapid amplitude changes}
\tablenotetext{d}{references for $T_{\mathrm{eff}}$: \citet{2008ApJ...688..377S}, \citet{2011A&A...527A..24L}}
\end{table*}

The measured periods range from 0.2 to 5\,d, with a median of 1.1\,d. The amplitudes of the lightcurves 
are typically a few percent, with the shape usually well approximated by a sinus curve. According to spectroscopy analysed
in the literature, the objects with measured periods have spectral types of M5 to M7.5 and effective temperatures
from 2500 to 3000\,K \citep{2008ApJ...688..377S,2011A&A...527A..24L}, corresponding to masses of 0.02 to 0.09$\,M{\odot}$
\citep{2015A&A...577A..42B}. The majority of these objects are likely at the high end of this mass range.
Given that our targets
are all mid to late M dwarfs and thus magnetically active objects, the most straightforward explanation for
the periodicities are magnetically induced cool spots on the surface, which modulate the flux over a rotational
cycle. Therefore, we interpret the periods as rotation periods. For a more detailed discussion on the causes
of periodic variability in very low mass objects, see \citet{2004A&A...419..249S,2005A&A...429.1007S}.

For the other 35 objects in the sample we cannot determine rotation rates. It is conceivable that the periods
measured here are biased, particularly since we are using an activity phenomenon (spots) to derive periods, and
activity may depend on rotation. Ultimately this can only be tested by complementary, activity-independent measurements of
rotation, such as spectroscopic $v\sin{i}$. At this point there is no evidence for a bias in the photometric period
samples; the rotation-activity relation is flat in mid M dwarfs and so far $v\sin{i}$ have confirmed the fast
rotation of very low mass objects \citep[e.g.][]{2012MNRAS.423.2966J}.
Therefore, for the remainder of the paper we will assume that these periods are representative of young BDs in UpSco. 

The periods are persistent throughout the campaign, i.e. over at least 72\,d, with stable amplitudes. 
Most objects show clear signs of phaseshift: If we plot the datapoints as function of phase for each of 
the segments, the maximum shifts through the observing campaign by 0.1-0.5 cycles over the course of 72\,d. 
Comparing the phased lightcurve for various segments also shows in some cases indications for changes in the
shape. Thus, while the surfaces of our targets are continuously covered by spots, the parts of the surfaces 
that induce the variability change over time. This is evidence for evolving spot pattern on
young BDs. In one case (EPIC204614722) the period significantly decreases over the observing 
campaign, from 0.69\,d to 0.61\,d, most likely because the variability is induced by spots located
at varying and differentially rotating latitudes. 

\begin{figure*}
\center
\includegraphics[width=7cm]{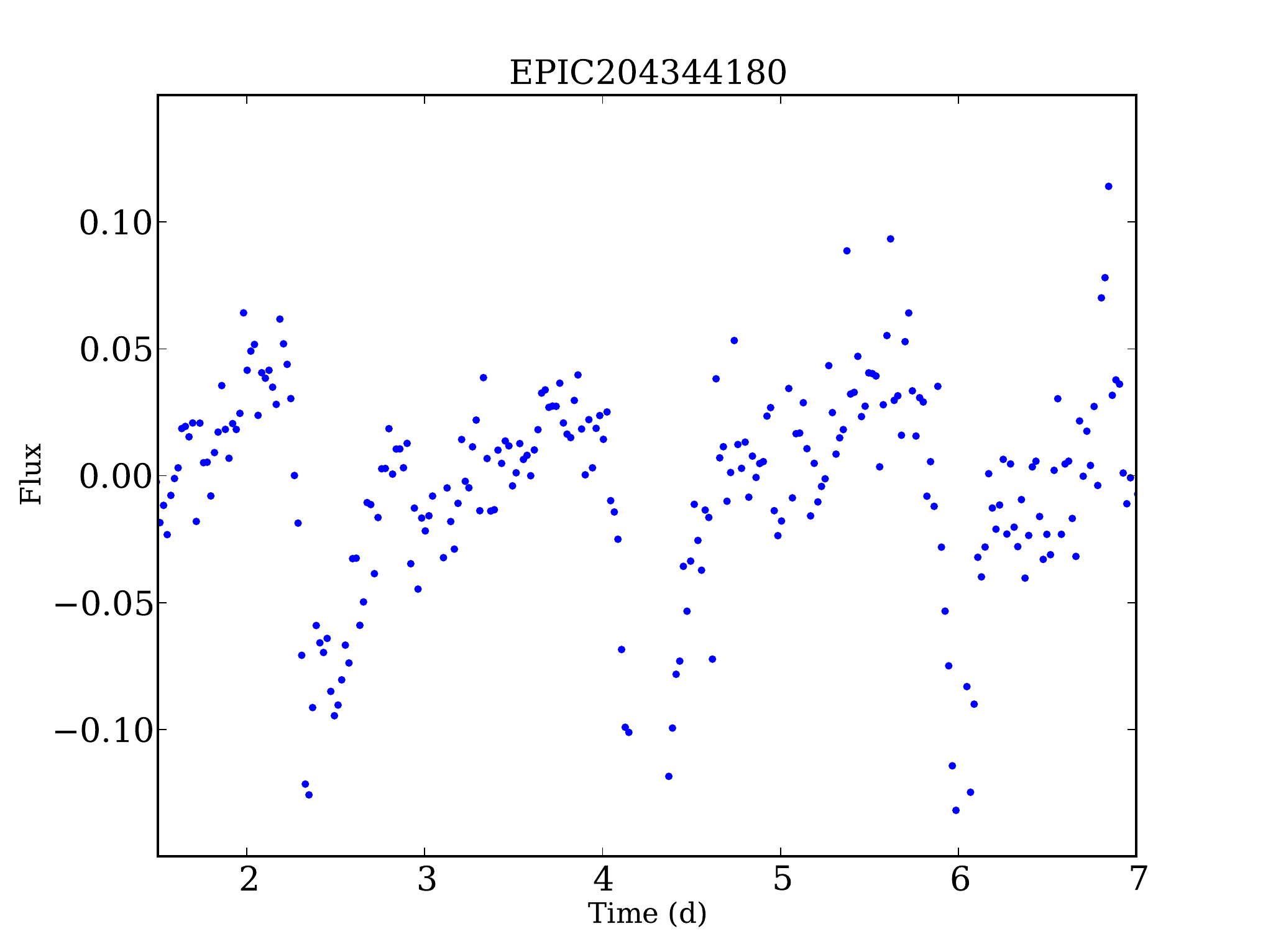}
\includegraphics[width=7cm]{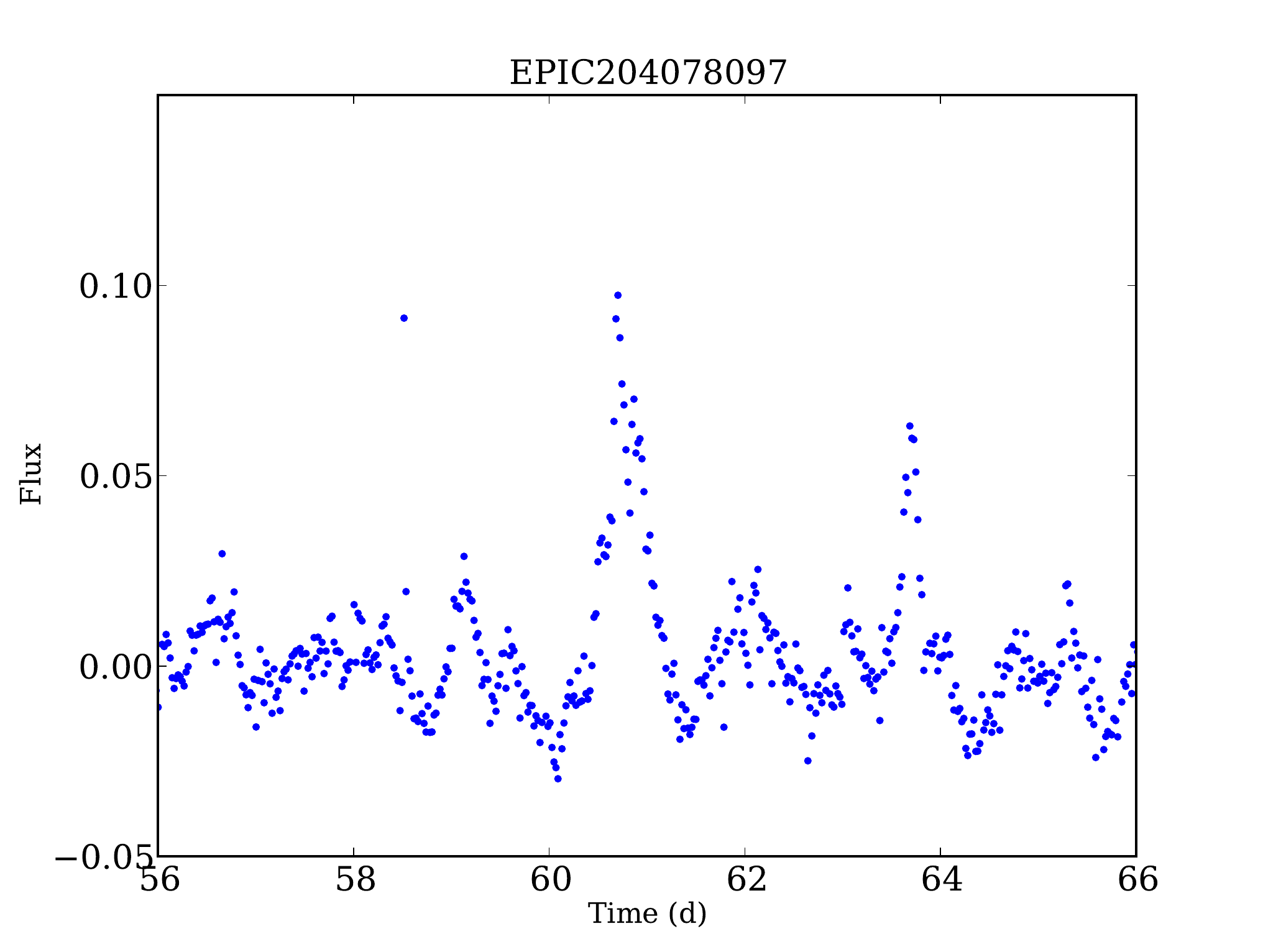}
\caption{Partial lightcurves for two objects with periodic variability to highlight the irregular eclipses (EPIC204344180,
left panel) and bursts (EPIC204078097, right panel).
\label{lc}}
\end{figure*}

Out of the 16 objects, 4 have mid-infrared colour excess indicating the presence of a disk \citep[Class II,][]{2013MNRAS.429..903D};
the remainder are disk-less based on their broadband colours (Class III). One of the Class II sources, EPIC204344180, shows 
a series of eclipses in the first part of its lightcurve (Fig. \ref{lc}, left panel). The 
width and depth of the features vary slightly, and the eclipses disappear after about 20\,d. The dominant periodicity, however, 
persists over the entire lightcurve. This object might belong to the object class of AA-Tau analogs, recently termed 'dippers', 
which show either regular or irregular eclipses, presumably caused by a wall at the inner edge of the disk 
\citep{2011ApJ...733...50M,2015AJ....149..130S}. To our knowledge, this is the first example of a 
BD 'dipper'. The fact that the rotation period of this object is consistent with the period of 
the eclipses indicates that the disk feature has to be near co-rotation radius. Another one of the sources with disk, 
EPIC204078097, has only insignificant colour excess at 3-5$\,\mu m$, but clearly exceeds the photospheric emission at 
12$\,\mu m$, which makes this a BD transition disk. This object shows irregular amplitude changes throughout the observing campaign
(Fig. \ref{lc}, right panel), possibly due to short accretion bursts, maybe analogous to the ones reported for stars by \citet{2014AJ....147...83S}
and \citet{2013ApJ...768...93F}. As pointed out above, EPIC203348744, another Class II object, 
features irregular variability which could at least partially be caused by accretion or the presence of a disk.

\section{Discussion}

The new periods for UpSco BDs give us an opportunity to probe the rotational evolution in the substellar 
regime. At ages of 1-20\,Myr, stars spin up when angular momentum is conserved, because they contract towards their 
final main-sequence radii. Angular momentum losses occur either due to magnetic star-disk coupling or accretion 
powered stellar winds \citep{2010ApJ...714..989M,2012ApJ...745..101M}. For a review of these processes and the 
observational evidence for disk braking, see \citet{2014prpl.conf..433B}

\begin{figure}
\center
\includegraphics[width=8cm]{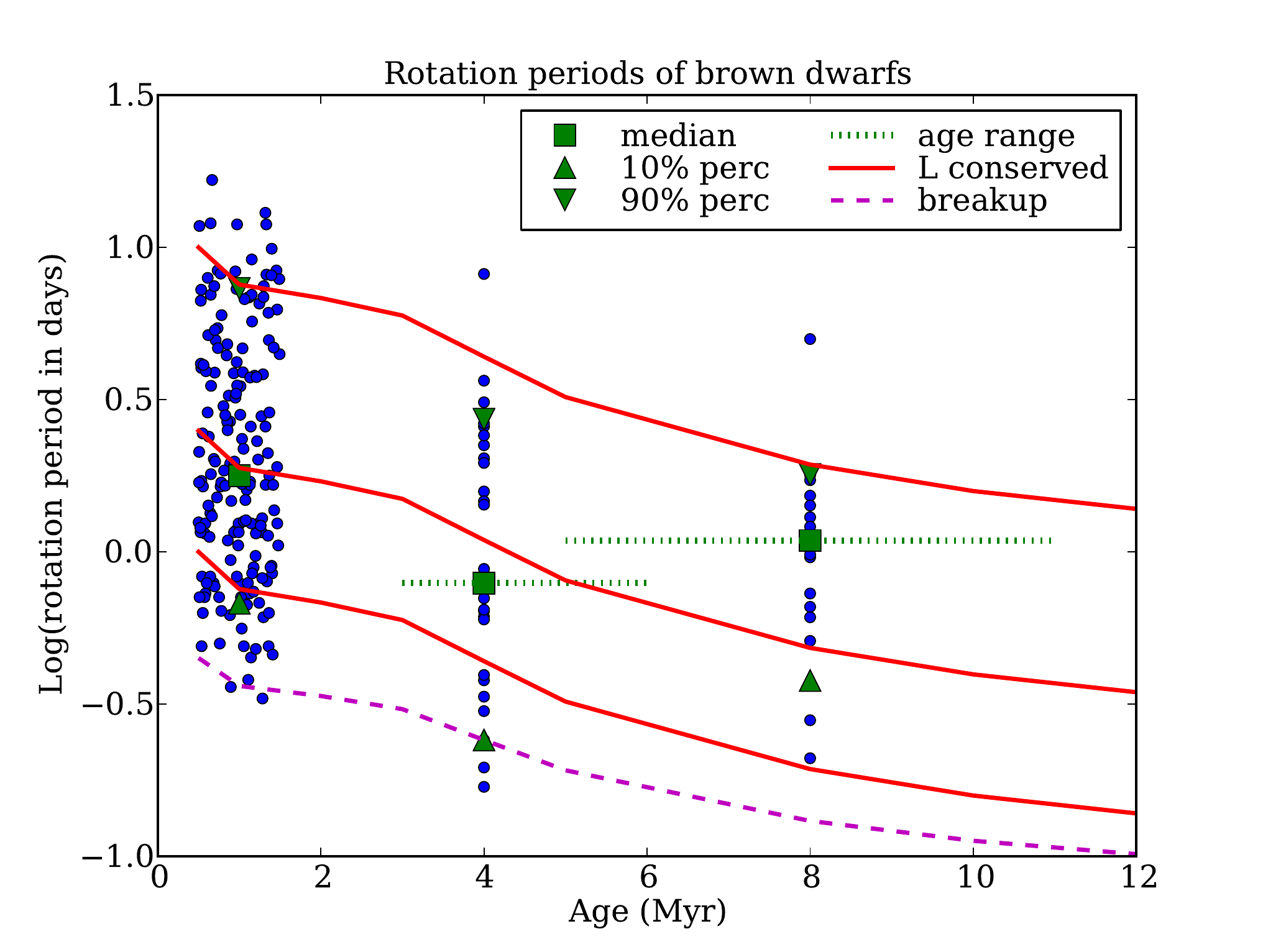} 
\caption{Rotation periods for brown dwarfs as a function of age, compared with evolutionary tracks assuming
angular momentum conservation. The breakup period is plotted as well. The periods at 1\,Myr (ONC) are randomly
spread out in age for clarity.
\label{rotevo}}
\end{figure}

In Fig. \ref{rotevo} we compare the rotation periods in UpSco presented in this paper (Table \ref{t1}) with the two other sizable 
samples of BD periods, in the ONC \citep{2009A&A...502..883R} and $\sigma$\,Ori\footnote{\citet{2003ApJ...594..971J} published 
3 more periods for young BDs in Chamaeleon-I (age 1-2\,Myr), with $P=2.2\ldots 3.4$\,d.} and with simple 
rotational evolution tracks. The models are calculated based on the evolutionary models by \citet{2015A&A...577A..42B}, 
for a fiducial BD mass of 0.07$\,M_{\odot}$. Adopting a different mass (0.04-0.07$\,M_{\odot}$) shifts 
the tracks slightly. All tracks assume angular momentum conservation. We also overplot the physical barrier for 
the rotation period ('breakup period').

For $\sigma$\,Ori, we combine the samples by \citet{2004A&A...419..249S} and \citet{2010ApJS..191..389C}. We also 
include in the $\sigma$\,Ori sample the BD periods for the cluster around $\epsilon$\,Ori \citep{2005A&A...429.1007S}, 
a poorly characterised region, but note that their inclusion does not change any of the following results. 
For the samples by \citet{2009A&A...502..883R} and \citet{2010ApJS..191..389C} we adopt a magnitude cutoff of $I=16.5$ 
to separate BDs from stars; this is subject to uncertainties in evolutionary tracks and cluster distances. 
For the other two samples, masses have been determined by comparing the IJHK photometry with evolutionary tracks
in the original papers. In total we use 179 periods in the ONC and 29 periods in $\sigma$\,Ori. We overplot the median 
and the 10\% and 90\% percentile. All samples should be sensitive to periods ranging from $\sim 0.2$ to several days, 
in the case of the K2 sample up to several weeks.

The choice for the ages of $\sigma$\,Ori and UpSco is relevant for the discussion. For $\sigma$\,Ori members the 
plausible age range is 3-6\,Myr \citep{2008AJ....135.1616S,2013MNRAS.434..806B}, for UpSco 5-12\,Myr 
\citep{2002AJ....124..404P,2012ApJ...746..154P}. Both regions are significantly older than the ONC, and UpSco is 
probably again significantly older than $\sigma$\,Ori. In Fig. \ref{rotevo} we choose ages of 4 and 8\,Myr for 
$\sigma$\,Ori and UpSco, respectively. Age spread is probably present in both regions, but since ages for individual 
BDs are not easily determined, this is neglected here.

The median and percentiles for the UpSco periods are slightly lower than in the ONC, consistent with spinup due to
contraction. Compared with $\sigma$\,Ori, the median and lower limit are actually slightly higher in UpSco. We do 
not put too much emphasis on this finding at this point, since we cannot be sure that we are not missing some of the 
ultrashort periods in the K2 lightcurves, which may have been removed together with systematics. The upper limit in
UpSco is lower than in $\sigma$\,Ori, mostly because $\sigma$\,Ori has a larger fraction of slow rotators with 
$P>2\,d$ (8/29 vs. 1/16). Some of the periods in $\sigma$\,Ori and $\epsilon$\,Ori are below or very close to the 
breakup period, this may have consequences on the rotational evolution, see discussion in \citet{2005A&A...429.1007S}.

From Fig. \ref{rotevo} it is evident that the current period census does not require any disk braking to explain 
the evolution of the period median and upper/lower limits. Only ages at the upper end of the plausible range 
(6\,Myr for $\sigma$\,Ori, 11\,Myr for UpSco) may require to include some disk locking in the evolutionary tracks, 
but only for the slow rotators and only over a locking timescale of $<2-3$\,Myr. This is about about half 
as long as the locking timescale for slowly rotating low-mass stars \citep{2005ApJ...633..967H,2013A&A...556A..36G}. 
Thus, independent of the choice of the ages, this analysis robustly demonstrates that disk braking is inefficient in 
the BD regime. This is consistent with earlier findings of 'moderate angular momentum loss' in very low mass stars,
by \citet{2005A&A...430.1005L}. It is notable that the four objects with disk in the UpSco sample are among the six 
slowest rotators, with periods of 1.4, 1.7, 1.8\,d, and 5\,d well above the median. This indicates that disk braking 
does have some effect, but only for very few objects and/or on short timescales.

\section{Summary}

We have measured a sample of rotation periods for young brown dwarfs in the UpSco region, using lightcurves from campaign
2 of the K2 mission. Our periods range from a few hours up to 5\,d, with a median just above one day, confirming that
BDs, except at the youngest ages, are fast rotators. Four of the slowest rotators have mid-infrared excess
emission due to the presence of a disk; at least two of them show signs of disk eclipses and accretion in the lightcurves. We compare
the new periods with previously published samples in the ONC and $\sigma$\,Ori and show that the period evolution from 
1 to 10\,Myr is consistent with no or little rotational braking, in contrast to low-mass stars. This confirms that 
disk braking, while still at work, is inefficient in the BD regime. This main finding is potentially 
an important constraint on the mass dependence of the braking mechanism. Compared with low-mass stars, young BDs 
have much lower accretion rates \citep[e.g.][]{2004A&A...424..603N}, weak magnetic fields \citep{2009ApJ...697..373R}, 
and possibly lower ionisation rates at the inner edge of the gas disk. These characteristics may influence
the efficiency of rotational braking.

\acknowledgements

We thank Thomas Barclay for his assistance with PyKE and Andrew
Vanderburg for a fruitful discussion on the details of his data reduction
technique. This work was supported in part by NSERC grants to R. Jayawardhana.
This work made use of PyKE \citep{2012ascl.soft08004S}, a software package for the reduction and 
analysis of Kepler data. This open source software project is developed and 
distributed by the NASA Kepler Guest Observer Office.

\end{document}